\documentstyle[twoside,fleqn,espcrc2]{article}

\newcommand{\AmS}{{\protect\the\textfont2
  A\kern-.1667em\lower.5ex\hbox{M}\kern-.125emS}}

\title{Self-organized criticality in atmospheric cascades}

\author{M.Rybczy\'nski, Z.W\l odarczyk
        \address{Institute of Physics, Pedagogical 
        University, Kielce, Poland \\
        emails: mryb@pu.kielce.pl and wlod@pu.kielce.pl}
        and 
        G. Wilk\address{The Andrzej Soltan Institute for Nuclear Studies, 
        Nuclear Theory Department, Warsaw, Poland\\
        email: wilk@fuw.edu.pl}}
       
\begin{document}

\begin{abstract}
We argue that atmospheric cascades can be regarded as example of the
self-organized criticality and studied  by using L\'evy flights and
nonextensive approach. It allows us to understand the scale-invariant
energy fluctuations inside cascades in a natural way.

\end{abstract}

\maketitle

\section{INTRODUCTION}

It is well known that energy spectra of particles from atmospheric
family events which are observed by the emulsion chambers at mountain
altitudes are essentially following power-like dependence. On the
other hand the occurence of power-law distributions is a very common
feature in nature and it is usually connected with such notions as
criticality, fractals and chaotic dynamics and studied using
generalized (nonextensive) maximum entropy formalism characterized by
a nonextensivity parameter $q$. Such formalism leads in a natural way
to power-laws in the frame of equilibrium processes \cite{T}. For
non-equilibrium phenomena the sources of power-law distributions are
self-organized criticality \cite{SOC} and stochastic multiplicative
processes \cite{SMP}. In the former nonequilibrium systems are
continuously driven by their own internal dynamic to a critical state
with power-laws omnipresent. In the later power-law is generated by
the presence of underlying replication of events. All these
approaches can be unified in terms of generalized nonextensive
statistics mentioned above (widely known as Tsallis statistics)
\cite{T}.\\ 

We shall look therefore at the development of cascades from this
point of view. To be more specific, we shall use the notion of L\'evy
flights as representatives of power-laws emerging from generalized
statistics \cite{T}.

\section{MONTE-CARLO SIMULATION OF MULIPLICATIVE PROCESSES}

\begin{figure}[h]
\setlength{\unitlength}{1cm}
\begin{picture}(7.28,7.11)
\includegraphics{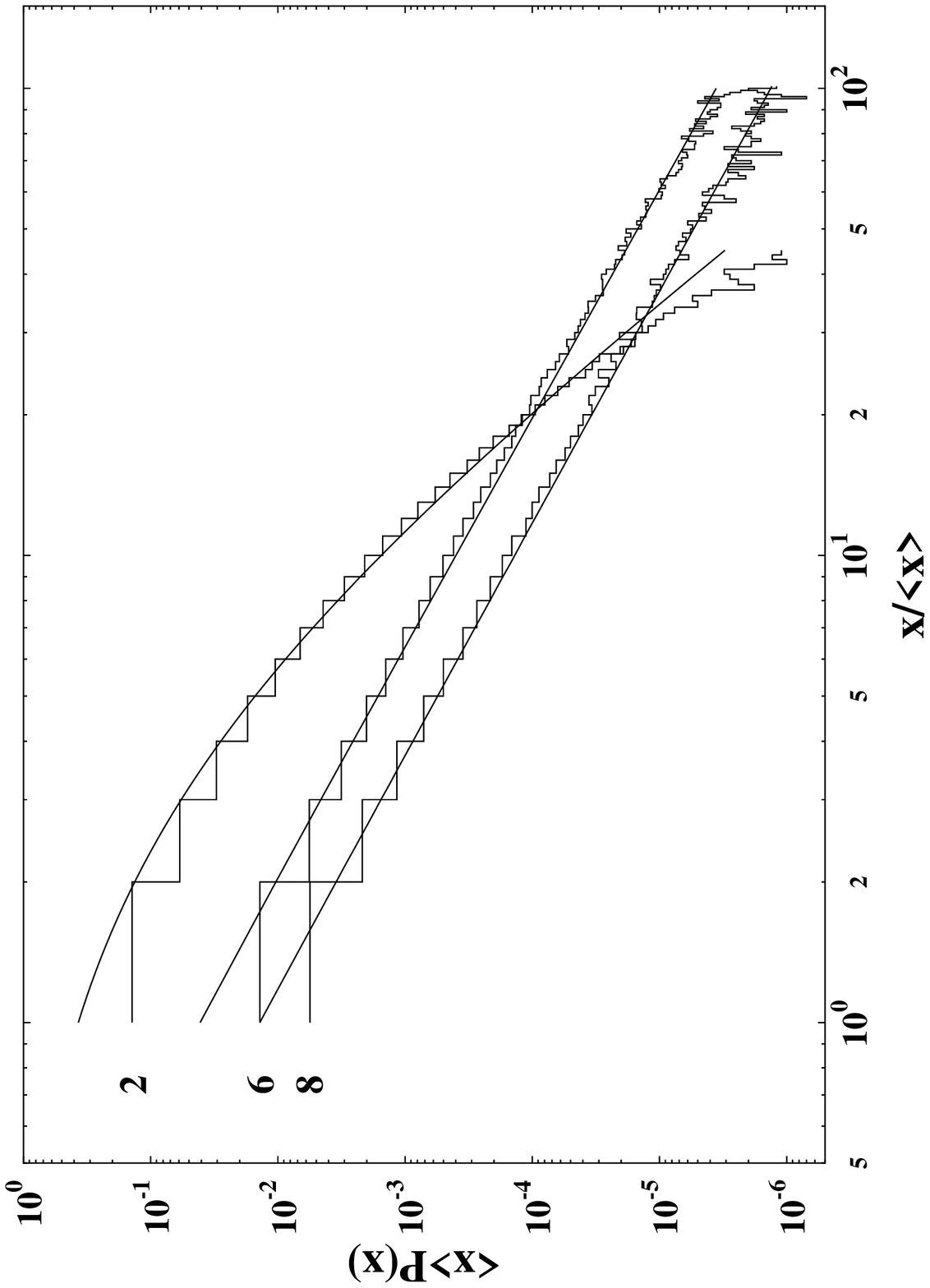}
\end{picture}
\vspace{-0.5cm}
\end{figure}
\vspace{-0.5cm}
\begin{minipage}[h]{6.8cm}
\noindent
Fig.1 Distributions $P(x)$ for multiplicative process, as given by eq.
(\ref{eq:MULTI}) (histograms), for different generation numbers 
$N=2, 6$ and $8$ are fitted by L\'evy distribution (eq. (\ref{eq:LEVY}))
(solid lines) with parameters $\alpha=4.81, 2.05$ and $2.015$,
respectively. 
\end{minipage}
\vspace{0.5cm}

Let us start with numerical (Monte-Carlo) consideration of the
following model of multiplicative process:
\begin{equation}
x_{N+1}\, =\, \xi\, x_N ,\qquad x_0=1, \label{eq:MULTI}
\end{equation}
where $\xi$ is random number taken from the exponential distribution
(to account for the essentially exponential scaled energy dependence
of the single elementary interaction):
\begin{equation}
f(\xi)\, =\, \frac{1}{\xi_0}\, \exp\left(- \frac{\xi}{\xi_0}\right). 
\label{eq:FX}
\end{equation}
As was shown in \cite{WW}, every exponential distribution with a
fluctuating parameter results in the L\'evy distribution (cf. Fig.
1), which in our case has the following form
\begin{equation}
P(x_N)\, =\, \frac{1}{\langle x_N\rangle}\cdot
             \frac{\alpha - 1}{\alpha - 2}\cdot 
             \left[1\, +\, \frac{1}{\alpha - 2}\frac{x_N}{\langle x_N\rangle} 
             \right]^{-\alpha}. \label{eq:LEVY}
\end{equation}             

\begin{figure}[h]
\setlength{\unitlength}{1cm}
\begin{picture}(7.28,7.11)
\includegraphics{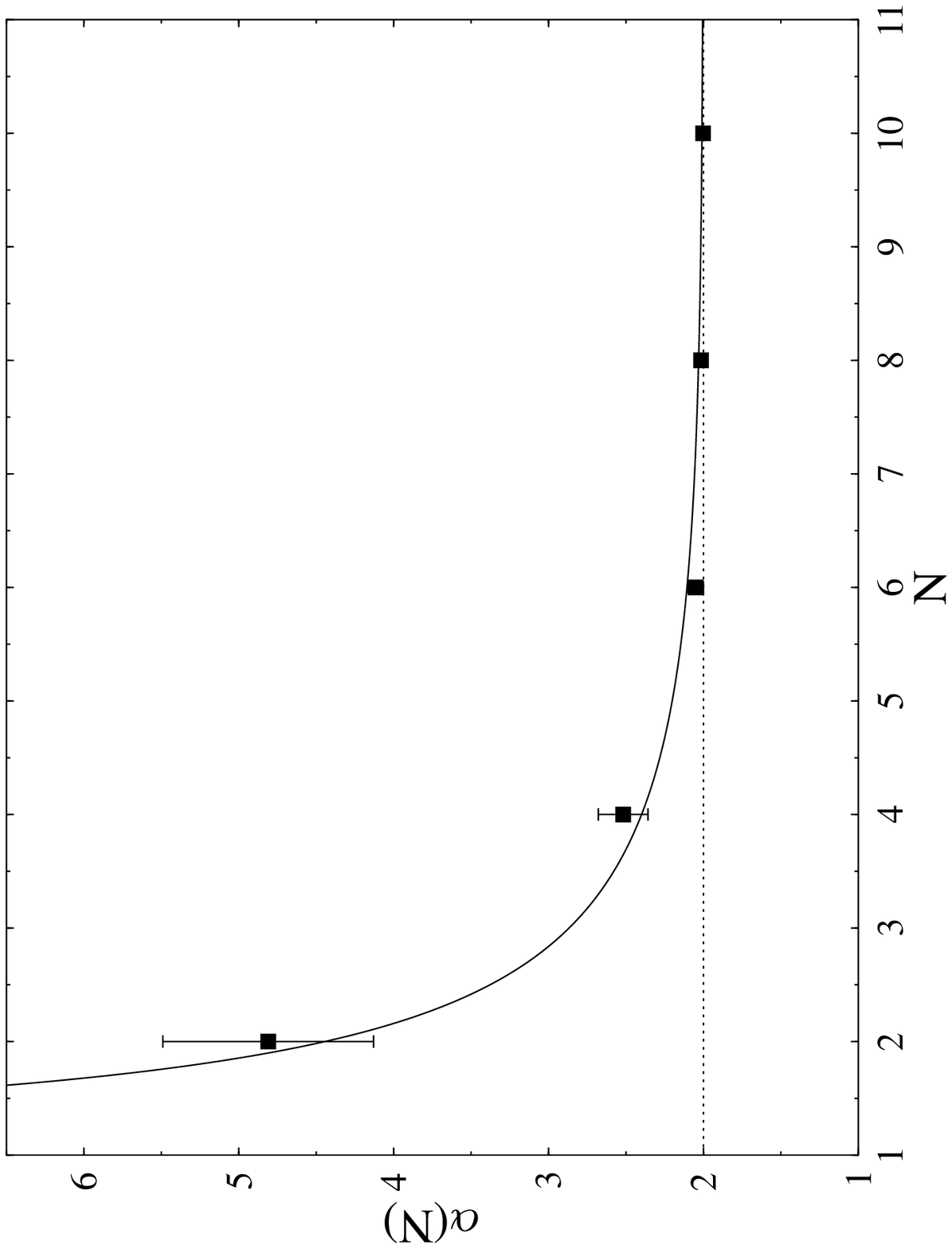}
\end{picture}
\vspace{-0.5cm}
\end{figure}
\vspace{-0.5cm}
\begin{minipage}[h]{6.8cm}
\noindent
Fig.2 Dependence of the parameter $\alpha$ of the Levy distribution
(\ref{eq:LEVY}) on the generation number $N$. Solid line represents
fit by the formula (\ref{eq:PAR}) with $c=0.55$. Asymptotically
$\alpha = 2$ (dotted line). 
\end{minipage}
\vspace{0.5cm}

In our present case parameter $\alpha$ depends on the number of
generations $N$ and on the parameter $\xi_0$ which defines
exponential distribution of the variable $\xi$ (in the limit of large
$N$ dependence on $\xi_0$ effectively vanishes). As is illustrated in
Fig. 2, dependence $\alpha(N)$ can be described by simple formula
\begin{equation}
\alpha\, =\, \alpha(N)\, =\, \frac{2}{1 - c^{N-1}} \label{eq:PAR}
\end{equation}
where $c$ is generation independent parameter which should be fitted
(in our case $c=0.55$). 

If we use the canonical form of L\'evy distribution as emerging from
Tsallis statistics \cite{T},
\begin{equation}
P(x)\, =\, \frac{2-q}{\chi}\left[1\, -\, (1 - q) \frac{x}{\chi}\right]
        ^{\frac{1}{1-q}} , \label{eq:LEVT}
\end{equation}
where mean $\langle x\rangle$ is given by the parameter $\chi =
(3-2q)\langle x\rangle$ and $q$ is mentioned before nonextensive
parameter such that $q=1+1/\alpha$ \cite{WW} then
\begin{equation}
q\, =\, \frac{3}{2}\, -\, \frac{c^{N-1}}{2}. \label{eq:Q}
\end{equation}
Notice that for $N\rightarrow 1$ parameter $q\rightarrow 1$ (or,
respectively, $\alpha \rightarrow \infty$), i.e., we
are recovering the initial exponential distribution. In the limit
$N\rightarrow \infty$ parameter $q$ approaches value $q=3/2$, which
is limiting value available for $q$ emerging from the normalization
condition imposed on the probability distribution $P(x_N)$. In this
limit $\alpha \rightarrow 2$. It is interesting to note that for
$[x_N/\langle x_N\rangle]\cdot [1/(\alpha - 2)] >> 1$,
i.e., for $x$ sufficiently large  one gets power-like behaviour of
$P(x_N)$: 
\begin{equation}
P(x_N) \, \propto\, \left(\frac{x_N}{\langle
x_N\rangle}\right)^{-\alpha} . \label{eq:INF}
\end{equation}
Actually such situation is reached reasonably fast, because $\langle
x_N\rangle = \xi^N_0$ and $\alpha$ tends to its limiting value
$\alpha=2$ rather quickly with increasing number of generations $N$.
In practice the equilibrium distribution $P(x_N)\, \propto \,
x_N^{-2}$ is reached (for $x_N >> \langle x_N\rangle$) already for 
$N>6$.

\section{COMPARISON WITH EXPE\-RI\-MEN\-TAL DATA}

There exists a number of experimental data from the emulsion chambers
exposed at mountain altitudes, which are relevant for our approach and which
we shall compare with now \cite{K,P}. Although, as we have already
mentioned, the energy spectra of particles from atmospheric family
events they represent are roughly expressed by the power-law
distributions, so far there was no analysis showing wheather it is
true and to what extend. Our work is the first attempt in this
direction as we show that they all can be described by the L\'evy type
spectra. 

\begin{figure}[h]
\setlength{\unitlength}{1cm}
\begin{picture}(7.28,7.11)
\includegraphics{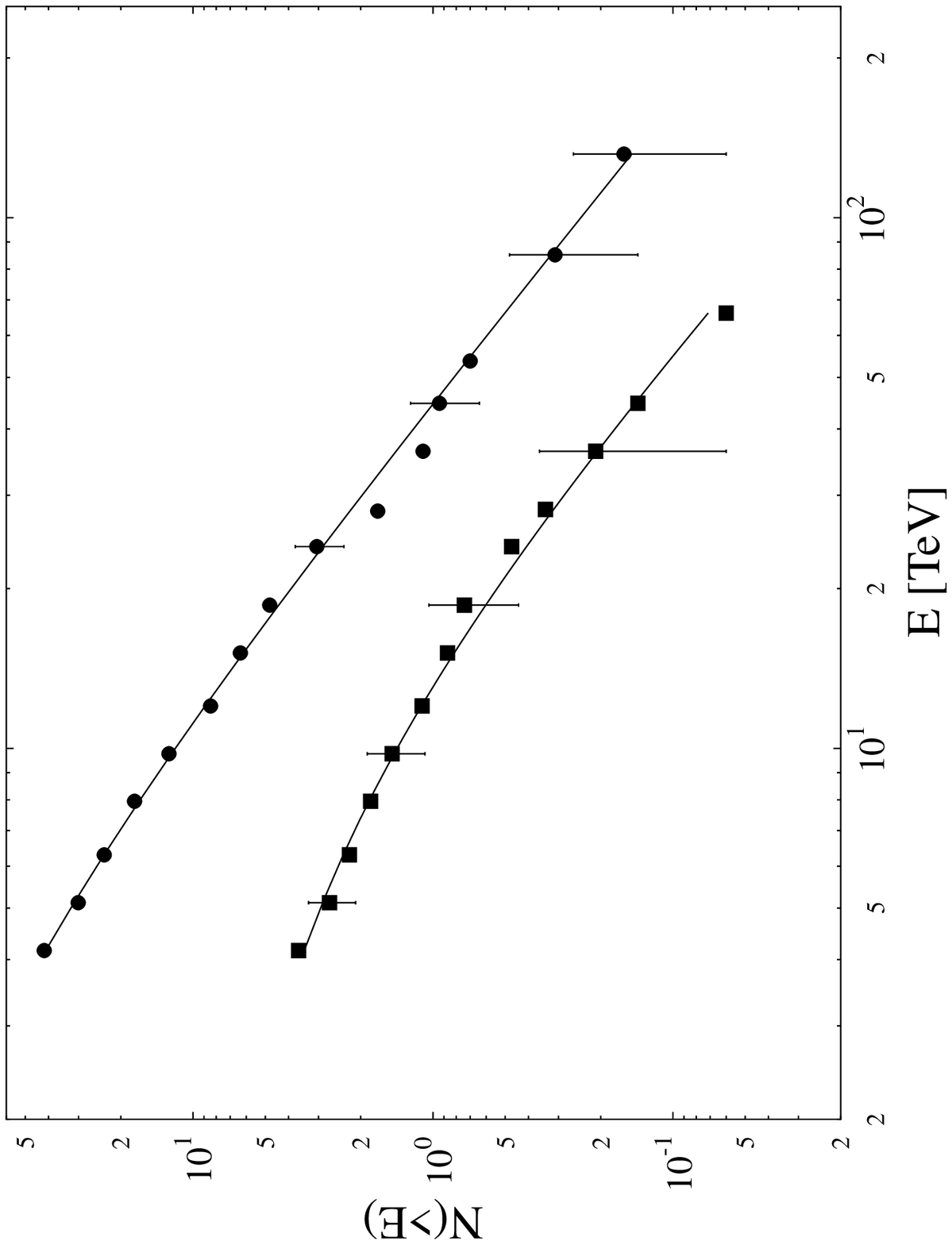}
\end{picture}
\vspace{-0.5cm}
\end{figure}
\vspace{-0.5cm}
\begin{minipage}[h]{6.8cm}
\noindent
Fig.3 Integral energy spectra for gamma quanta (circles) and hadrons
(squares) in families registered at Mt. Kambala are fitted by L\'evy
distributions (solid lines).
\end{minipage}
\vspace{0.5cm}

Integral spectra $N(>E)$ from Kambala experiment (at $520~{\rm g}/{\rm
cm}^2$) \cite{K} are shown in Fig. 3 where they are fitted using
parameters $\alpha -1 = 1.8$ (for gamma families) and $\alpha-1=2.0$
(for hadronic families). Notice that hadronic component is little
"younger" (higher $\alpha$ means smaller $N$, cf. Fig. 2) than the
electromagnetic component.

Families observed deeper in the atmosphere (what means higher $N$)
have smaller $\alpha$, as expected, cf. Fig. 4, where $N(>E)$ from
Pamir experiment (at $600~{\rm g}/{\rm cm}^2$) \cite{P} are presented.
Here the corresponding parameters $\alpha - 1$ are equal to $1.8$ and
$1.7$ for gamma and hadronic components, respectively.

It is evident from Figs. 3 and 4 that the observed gamma-hadron
families are already reaching their quasi-equilibrium states. It
should be noticed that most of the families registered come not from
a single nuclear interaction but they usually contain particles which
are decascadents of several ($\sim 7$ at $\Sigma\, E > 100$
TeV) genetically connected nuclear interactions \cite{NTW}. 
Our analysis indicates 
that the average number of cascade generations leading to the
observed distributions is about $N\simeq3$.

\begin{figure}[h]
\setlength{\unitlength}{1cm}
\begin{picture}(7.28,7.11)
\includegraphics{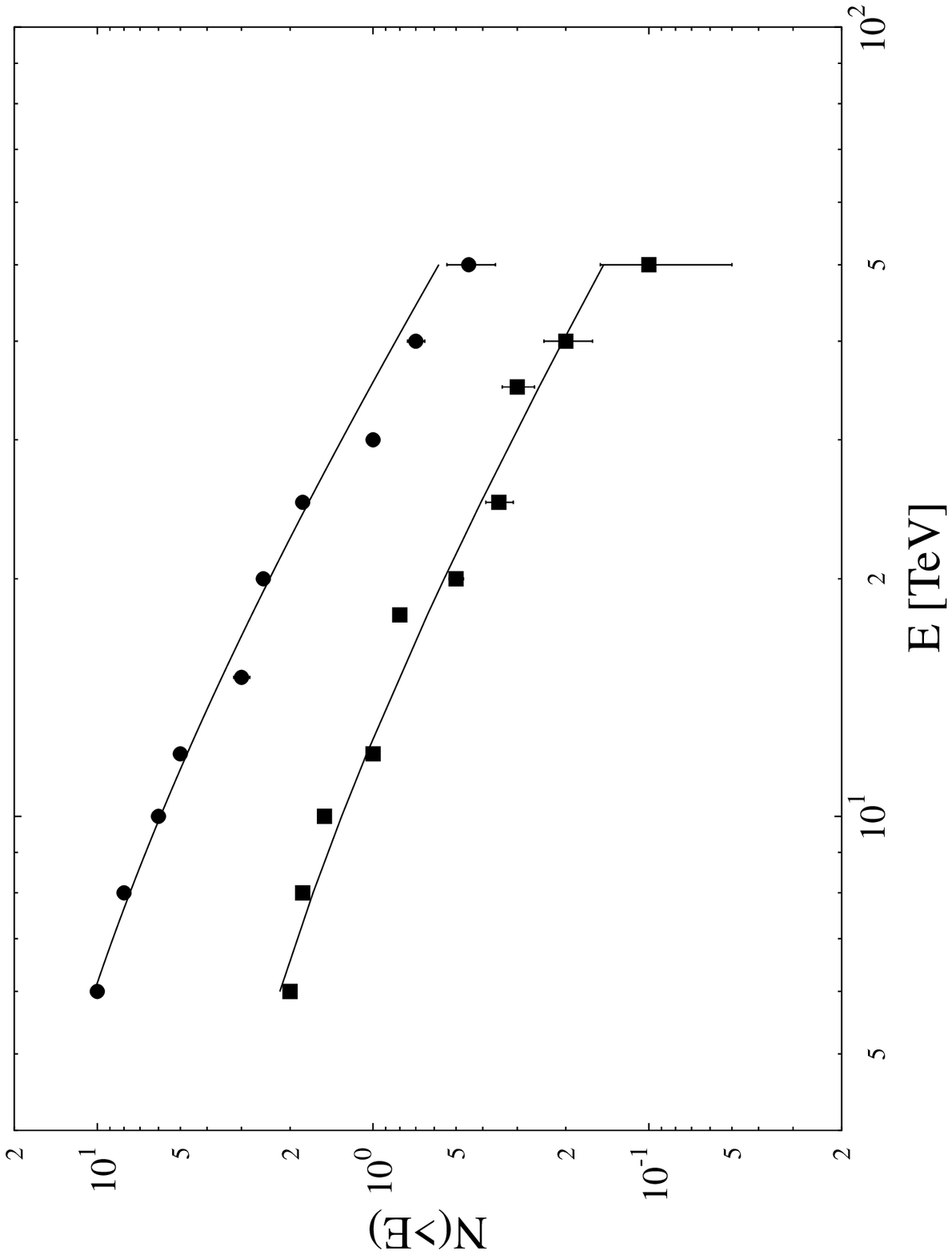}
\end{picture}
\vspace{-0.5cm}
\end{figure}
\vspace{-0.5cm}
\begin{minipage}[h]{6.8cm}
\noindent
Fig.4 Integral energy spectra for gamma quanta (circles) and hadrons
(squares) in families registered at Mt. Pamir are fitted by L\'evy
distributions (solid lines).
\end{minipage}
\vspace{0.5cm}

\section{SUMMARY AND DISCUSION}

We have provided description of cascade processes encountered in
cosmic ray emulsion experiments at high altitudes, which is based on
the notion of L\'evy distrubutions (which, in turn, originate from
the nonextensive statistics described by parameter $q$ \cite{T,WW}).
In this way we have found that the observed distributions are to a
high accuracy power-like, but at the same time we are able to account
for the small deviations from the exact power-like behaviour. They
are, in our approach, directly connected with the finite number $N$
of the cascade generations in the way provided by formulas
(\ref{eq:PAR}) or (\ref{eq:Q}). 

The results presented before should be confronted with the result
of pure nuclear origin, which is the case of the so called "halo"
events with strong concentration of particles in the central region.
Fig. 5 shows such a case exposing integral energy distribution of
shower cores (recorded at distance $R<0.65$ cm from the
energy-weighted center) obtained in a "halo" event P06 registered at
Chacaltaya \cite{C}. In this case we observe almost exponential
distribution ($\alpha - 1 = 16.0$ in this case, what translates to
$q=1.06$) with no influence of multiplicative processes.

\begin{figure}[h]
\setlength{\unitlength}{1cm}
\begin{picture}(7.28,7.11)
\includegraphics{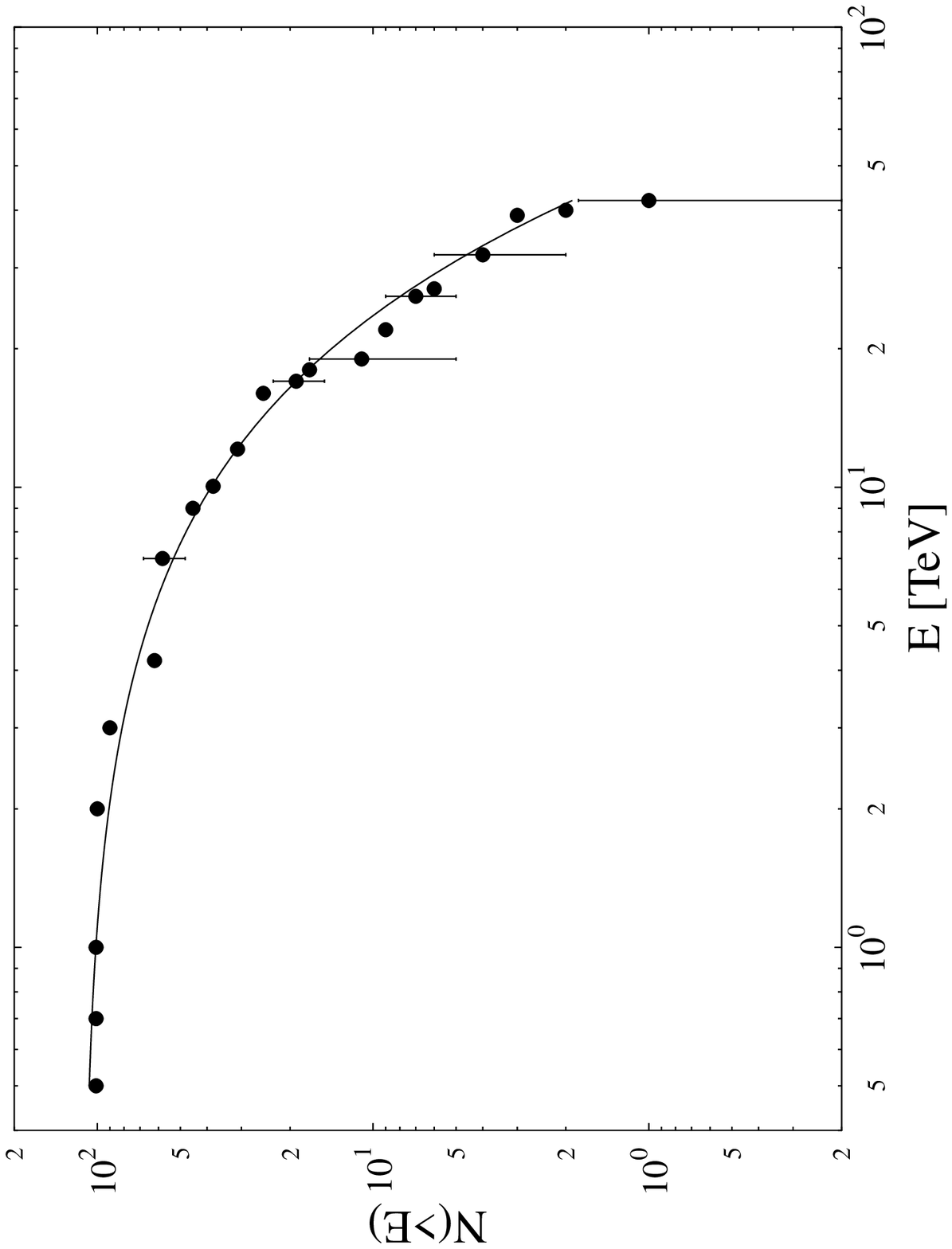}
\end{picture}
\vspace{-0.5cm}
\end{figure}
\vspace{-0.5cm}
\begin{minipage}[h]{6.8cm}
\noindent
Fig.5 Integral energy spectra for shower cores from halo event (circles) 
registered at Chacaltaya is fitted by L\'evy distribution (solid lines).
\end{minipage}
\vspace{0.5cm}

Finally, it should be noticed that there exists a number of special cases
more-or-less directly connected with our results:\\
$(a)$ Energy transport equation \cite{ETE} for cascade processes gives
\begin{equation}
N(E,t)\, =\, N_0 E^{-(s+1)}\exp\left(-\frac{s-1}{s+1}\cdot\frac{t}{t_0}\right)
\end{equation}
as number of particles with energy $E$ at depth $t$. Notice that
for the age parameter $s\rightarrow 1$ one approaches equilibrium
spectrum $N(E)=N_0\, E^{-2}$ with the slope $\alpha = 2$.\\
$(b)$ One observes strong intermittent behaviour in the families
\cite{INT}, which results from the fluctuations in the atmospheric
cascades alone and is not sensitive to details of the elementary interactions.\\
$(c)$ One observes multifractal behaviour in the cascades with
fractal dimension $D_N$ at each cascade stage $N$ where $P(x) \propto
x^{D_N}$ \cite{FRAC}. For each successive emission the distribution
should be more inhomoheneous, i.e., $D_N > D_{N+1}$.\\
$(d)$ Self-organized criticality \cite{SOC} to which all cascades
considered here bear close resemblance because of the Zipf's law \cite{Z},
which predicts power-law behaviour of the type seen in eq.(\ref{eq:INF})
(with $\alpha =2$).\\

Acknowledgements: The partial support of Polish Committee for 
Scientific Research (grants 2P03B 011 18 and 
621/E-78/SPUB/CERN/P-03/DZ4/99) is acknowledged.\\


\begin{thebibliography}{9}

\bibitem{T} See, for example, G.Wilk and Z.W\l odarczyk, {\sl Nucl.
            Phys.} {\bf B} (Proc. Suppl.) {\bf A75} (1999) 191 and
            references therein. Cf. {\sl Braz. J. Phys.} {\bf 29} No.1 (1999)
            (also at the URL:
            http://sbf.if.usp.br/ WWW\_pages/Journals/ BJP/Vol29/Num1/
            index.htm) for review.
            
\bibitem{SOC} P.Bak, C.Tang and K.Wiesenfeld, {\sl Phys. Rev. Lett.}
              {\bf 59} (1987) 381; M.Paczuski, S.Maslow and B.Pak,
              {\sl Phys. Rev.} {\bf E53} (1996) 414.  

\bibitem{SMP} S.C.Manrubia and D.H.Zanette, {\sl Phys. Rev.} {\bf
              E59} (1999) 4945.

\bibitem{WW} G.Wilk and Z.W\l odarczyk, {\sl Phys. Rev. Lett.} {\bf
             84} (2000) 2770.

\bibitem{K} J.R.Ren et al., (Kambala Coll.), {\it Proc. $19^{th}$
            Int. Cosmic Ray Conf.}, La Jolla, {\bf 6} (1985) 429.

\bibitem{P} A.S.Borisov et al. (Pamir Coll.), {\sl Bull. Soc. Sci.
            Lett. de \L \'od\'z}, {\bf XLII-5} (1992) 71.
            
\bibitem{NTW} J.Nowicka, A.Tomaszewski and Z.W\l o\-dar\-czyk, {\sl J.
              Phys.} {\bf G11} (1985) 1365.

\bibitem{C} N.M.Amato, N.Arata and R.H.C.Maldonato, {\t Proc. Int.
            Symp. Cosmic Ray Superhigh Energy Int.}, Beijing (1986)
            4-37.             

\bibitem{ETE} S.Hayakawa, {\it Cosmic Ray Physics}, John Willey \&
              Sons, New York, 1969.

\bibitem{INT} G.Wilk and Z.W\l odarczyk, {\sl J. Phys.} {\bf G19}
              (1993) 761.

\bibitem{FRAC} I.Sarcevic and H.Satz, {\sl Phys. Lett.} {\bf B233}
               (1989) 252; cf. also: S.V.Chekanov, {\sl Eur. Phys.
               J.} {\bf C6} (1999) 331.

\bibitem{Z} Cf., for example, R.V.Sole, S.C.Manrubia, M.J.Benton and
            P.Bak, {\sl Nature} {\bf 388} (1997) 764 and references
            therein. 

\end{thebibliography}
\end{document}